\documentclass[reprint,amsmath,amssymb,aps,prl,groupedaddress,showpacs]{revtex4-1}
\usepackage{latexsym,amscd,amsopn,amsthm,amsfonts}
\usepackage{color,graphicx}%
\usepackage{dcolumn}%
\usepackage{bm}%
\renewcommand{\Re}{\mathop{\rm Re}}
\renewcommand{\Im}{\mathop{\rm Im}}

\begin{document}
\title{Coherently controlling Raman-induced grating in atomic media}

\author{V. G. Arkhipkin$^{1,2}$}%
\email{avg@iph.krasn.ru}

\author{
S. A. Myslivets$^{1,3}$}%
\email{sam@iph.krasn.ru}%

\author{
I. V. Timofeev$^{1}$}%
\email{tiv@iph.krasn.ru}%

\affiliation{$^1$L.V.Kirensky  Institute of Physics,
Krasnoyarsk 660036, Russia \\
$^2$Laboratory of Nonlinear Optics and Spectroscopy and \\
$^3$Department of Photonics and Laser Technology,Siberian Federal University,
Krasnoyarsk 660079, Russia  }%
\date{}%


\begin{abstract}
We consider dynamically controllable periodic structures, called Raman induced gratings, in  three- and four-level atomic media, resulting from  Raman interaction in a standing-wave pump. These gratings are due to  periodic spatial modulation of the Raman nonlinearity and fundamentally differ from the ones based  on electromagnetically induced transparency.
The transmission and reflection spectra of such gratings  can be simultaneously amplified and controlled by varying the pump field intensity.
It is shown that a transparent medium with  periodic spatial modulation of the Raman gain can be opaque near the Raman resonance and yet at the same time it can be  a non-linear amplifying mirror.
We also show that spectral properties of the Raman induced grating can be controlled with the help of an additional weak control field.
\end{abstract}

\maketitle

\section{INTRODUCTION}

Nonlinear periodic optical media have important applications in controlling light propagation behavior, which has attracted a great deal of attention \cite{Denz:2009}. Spatially micromodulated materials such as photonic crystals (PCs)  have been studied both theoretically and experimentally for more then two decades \cite{Joann:book.PC2008}. A large majority of PCs are formed by periodic modulation of the refractive index, i.e.  modulation of  the real part of the dielectric constant. The presence of absorption and (or) gain can radically change the nature of wave propagation in such materials, including the very notion of forbidden and allowed bands \cite{Rozanov:JETP.114.782}.

Quite a distinct approach to PCs uses optical resonant or quasiresonant nonlinearities that can be induced in a multilevel medium. Such nonlinearities can be exploited to generate spatially periodic structures required  to devise an optically tunable photonic band gap (PBG).  Unlike ordinary PCs,  here a periodic structure is created by external control light beams. Novel photonic structures  could be created by optically inducing coherent nonlinearities based on electromagnetically induced transparency (EIT) \cite{Fleisch:RMP.77.633} in a standing-wave coupling configuration \cite{Wu:JMO.56.768}.

When a standing-wave coupling field interacts with a three-level atomic system, the dispersion and absorption of a probe laser beam in an atomic medium is modulated spatially by the standing-wave coupling field. It  has been proposed  to induce spatially periodic quantum  coherence  for  generation of a tunable PBG \cite{Andre:PRL.89.143602,Su:PRA.71.013821,Artoni:PRL.96.073905,Kuang:OE.16.15455,Wu:PRB.77.113106} and  dynamic generation of stationary light pulses \cite{Bajcsy:Nat.426.638,Hansen:PRA.75.053802,Moiseev:PRA.89.043802}.
These structures are also referred to as electromanetically induced absorption gratings (EIAG) \cite{Brown:OL.30.699} or  electromanetically induced  gratings (EIG) \cite{Ling:PRA.57.1338}.
Recently, many researchers  have been focused on the study of EIAG \cite{Ling:PRA.57.1338,Mitsunaga:PRA.59.4773,Zhai:PhysL.289.27, Cardoso:PRA.65.033803,Brown:OL.30.699,Dutta:JPB.39.1145, Bae:AO.47.4849, Araujo:OL.35.977,Xiao:JPB.43.161004,Wen:PRA.82.043814},
that are based on EIT, with their  potential applications in mind.
EIAG may be utilized for diffracting and switching a probe field, and probing the optical properties of materials.

The wave propagation in resonant optical media with an active-Raman-gain (ARG) core has also attracted considerable theoretical and experimental interest \cite{Wang:Nat.406.277,Dogariu:PRA.63.053806,Payne:PRA.64.031802, Jiang:PRA.74.041803,Jiang:PRA.76.033819}. Unlike the EIT-based scheme, which is inherently absorptive, the central idea of the ARG scheme is that the probe (Raman) field operates in a stimulated Raman emission mode, and hence can eliminate signal attenuation  and can help realize a stable subluminal (slow light) as well as superluminal  (fast light) propagation of the probe wave even at room temperature  \cite{Payne:PRA.64.031802, Jiang:PRA.74.041803,Jiang:PRA.76.033819}. Such systems could enable new practical applications of coherent processes.
Recently it has been shown  that a gain-assisted giant Kerr effect \cite{Deng:PRL.98.253902,Jiang:PRA.77.045804,Agarwal:PRA.70.023802} and superluminal solitons can be obtained using an ARG medium  \cite{Hang:OE.18.2952}. In papers \cite{Arkh:PRA.80.061802,Arkh:JETP.111.898,Arkh:PRA.88.033847} it was theoretically shown how one could use an  ARG medium together with a PC cavity to create all-optical switches and transistors.

Recently we proposed a new concept of electromagnetically induced gratings in  three-level atomic media, called Raman induced gratings (RIG),  on the basis of spatial modulation of the Raman nonlinearity in a standing-wave pump field \cite{Arkh:OL.39.3223}.
In this paper, we present a systematic investigation of this concept and propose a scheme to coherently control propagation of the probe (Raman) wave in a four-level atomic medium of N-type using an additional control field.
Such  extension opens new possibilities for controlling the spectral properties of  RIG. The transmission and reflection  spectra can be controlled by  varying the intensity or frequency of the control field.

\section{THEORETICAL MODEL AND BASIC EQUATONS}\label{model}

\begin{figure}
  \centering
   \includegraphics[width=.5\columnwidth]{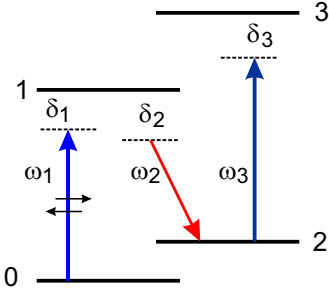}
     \caption{(Color online) Energy-level configuration and excitation scheme of a four-state ARG system of N-type, which interacts with a standing pump wave of frequency $\omega_1$, a weak traveling probe (Raman) field of frequency $\omega_2$ and a traveling control field of frequency $\omega_3$.  }\label{fig1}.
\end{figure}

The four-level scheme of $N$-type configuration intended for coherent control of the Raman gain process is shown   in Fig.~\ref{fig1}. We consider an ensemble of a lifetime broadening four-level atomic system initially prepared in the ground state $|0\rangle$. The transitions  $|0\rangle$-$|1\rangle$,  $|1\rangle$-$|2\rangle$ and $|2\rangle$-$|3\rangle$ are electric dipole allowed and   transition $|0\rangle$-$|2\rangle$ and $|1\rangle$-$|3\rangle$ are electric dipole forbidden.
The states $|0\rangle$ and $|1\rangle$ are coupled by a standing pump wave that is formed by two monochromatic counterpropagating pump fields $E_p=1/2\{E_1^+\exp[-i(\omega_1t-k_1z)]+E_1^-\exp[-i(\omega_1t+k_1z)]\}$, where $E_1^+$ and  $E_1^-$ are the amplitudes of the forward (FW) and backward (BW) wave, respectively, with frequency $\omega_1$ and wave vector $k_1$. For simplicity, we assume that these amplitudes are real.
A traveling weak probe Raman wave $E_s=1/2E_2\exp[-i(\omega_2t-k_2z)]$ and a control field $E_c=1/2E_3\exp[-i(\omega_3t-k_3z)]$  also propagate along the $z$ direction and interact with the transitions  $|1\rangle$-$|2\rangle$ and $|2\rangle$-$|3\rangle$, respectively. The pump  field is detuned from the state $|1\rangle$ with a large one-photon detuning so that single-photon absorption of the pump can be neglected. We will assume the probe field intensity to be much lower than the intensity of the pump field. The probe being weak, levels $|1\rangle$ and $|2\rangle$ remain empty and the space distribution of atoms remains homogeneous within the sample.

An off-resonant standing pump wave and the probe field induce spatially modulated coherence on the transition $|0\rangle$-$|2\rangle$ (off-diagonal element of the density matrix $\rho_{20}$), which leads to  modulation of the Raman nonlinearity susceptibility.
As a result, a medium with a spatially modulated  Raman gain and  nonlinear refractive index for the probe light is created.  Therefore, the weak probe wave propagates in such medium as  in a one-dimensional periodic structure with a period $\Lambda=\lambda_1/2$, where $\lambda_1$ is the wavelength of the pump field. The intensity of the pump  is chosen so as not to exceed the stimulated Raman scattering threshold while being sufficiently strong to ensure a noticeable enhancement of the probe wave. The control field is applied on the transition $|2\rangle$ and $|3\rangle$ and will enable us to manipulate the Raman gain.

The Raman gain coefficient and the nonlinear (cross-Kerr) refractive index  for the probe field  are determined by the imaginary and real part of the macroscopic nonlinear susceptibility $\chi_{nl}(\omega_2)$. Using the full density matrix equations and solving them exactly in the control field, to the second order in the pump field and the first order in the probe field, we find the nonlinear susceptibility $\chi_{nl}(\omega_2)$, that depends on the pump and control fields:

\begin{align}
\chi_{nl}(\omega_2)&=\frac{1}{2}\chi_R(\omega_2)N|E_p|^2\nonumber\\
\chi_R(\omega_2)&=\frac{D_{02} (\Delta_{31} \Delta_{30}+|\Omega_3|^2)}
{\Delta_{10}[\Delta_{20} \Delta_{30}-|\Omega_3|^2][\Delta_{12}^\ast \Delta_{31}+|\Omega_3| ^2]}, \label{eq1}
\end{align}
where  $E_{p}=E_1^+e^{ik_1z}+E_1^-e^{-ik_1z}$ with $2\Omega_1^+=d_{01}E_1^+/\hbar$ and $2\Omega_1^-=d_{01}E_1^-/\hbar$ representing the Rabi frequency amplitudes of the FW and BW plane-wave components  of the pump field,
$2\Omega_{3}=d_{23}E_{3}/\hbar$ is the Rabi frequency of the control field, $d_{ij}$ is the matrix dipole moment of the transition, and $\hbar$ is the reduced Planck constant,
$\Delta_{10} =\delta_1 +i\gamma_{10}$, $\Delta_{21} =\delta_2 +i\gamma_{12}$,  $\Delta_{20}=(\delta_1 -\delta_2) +i\gamma_{20}=\delta_{20}+i\gamma_{20}$, $\Delta _{30}=(\delta_1 -\delta_2+\delta_3) +i\gamma_{30}
=\delta_{30}+i\gamma_{30}$, $\Delta_{31}=(\delta_3-\delta_2)+i\gamma_{31}=\delta_{31}+i\gamma_{31}$, and $\delta_j=\omega_{mn}-\omega_j$ are the frequency detunings, $\delta_{20}=\omega_{20}-(\omega_{1}-\omega_{2})$, $\delta_{30}=\omega_{30}-(\omega_{1}-\omega_{2}+\omega_{3})$, $\delta_{31}=\omega_{31}-(\omega_{3}-\omega_{2})$, $\omega_{mn}$ and $\gamma_{mn}$ are the frequencies and halfwidths of the respective transitions, $D_{02}=|d_{01}|^2|d_{12}|^2 N/4\hbar^3$, $N$ is the concentration of atoms.
Hereinafter we assume that $|\delta_{1,2}|\gg\gamma_{10},\gamma_{12}$ and the Rabi frequency of of the pump field.

The Raman gain and refractive index for the probe wave can be described by the dielectric function
\begin{widetext}
\begin{equation}\label{WE}
\varepsilon_2(z)=1+4\pi N \chi_{l}(\omega_2)+4\pi N \chi_R(\omega_2)|E_{p}|^2=\varepsilon_{2l}+\Delta\varepsilon(1+p\cos{(2k_1z)}).
\end{equation}
\end{widetext}
Here $\varepsilon_{2l}=1+4\pi N\chi_{l}(\omega_2)$, $\chi_{l}(\omega_2)$ is the off-resonant macroscopic linear susceptibility for the probe field, $\Delta\varepsilon=4\pi N\chi_R(\omega_2)|E_p|^2$, $|E_p|^2=|E_1^+e^{ikz}+E_1^-e^{-ikz}|^2)=(E_1^{2+}+E_1^{2-})[1+p\cos(2k_1z)]$, $p=2E_1^+E_1^-/(E_1^{2+}+E_1^{2-})$.
Clearly, the modulation depth is maximal when $p=1$ when $E_1^+=E_1^-$.
It should be noted that $|\Delta\varepsilon|\ll 1$, which corresponds to a shallow depth of modulation.
When $\delta_{20}\neq 0$, a hybrid RIG grating is induced in the medium: a gain grating and a refraction
grating. The former is an amplitude grating, and  the later is a phase grating.

Let us consider normal incidence of a weak probe wave on a one-dimensional periodic medium with the permittivity \eqref{WE}. A sample of finite length $L$ contains a very large number of periods $\Lambda$. The wave equation for the electric field strength $E_2(z)$ inside a  medium with a spatially modulated dielectric constant  has the form \cite{Rautian:OS.104.112}
\begin{equation}\label{Eq:E2}
\frac{d^2E_2}{dz^2}+k_{20}^2\varepsilon_2(z)E_2=0,
\end{equation}
Considering Eq.~\eqref{WE}, equation \eqref{Eq:E2} can be reduced to
\begin{equation}\label{WE1}
\frac{d^2E_2}{dz^2}+k_2^2[1+\mu(1+p\cos{(2k_1z)})]E_2=0,
\end{equation}
where $k_{20}=\omega_2/c$, $k_2^2=k_{20}^2\varepsilon_{2l}$, $\mu=\Delta\varepsilon/\varepsilon_{2l}$ and $c$ is the velocity of light in vacuum.

By using the method of coupled waves \cite{Rautian:OS.104.112}, solution of Eq.~\eqref{WE1} can be represented as a superposition of two waves propagating in opposite directions.
\begin{equation}\label{E2}
E_2(z)=A(z)e^{ik_2z}+B(z)e^{-ik_2z},
\end{equation}
where $A(z)$ and $B(z)$ are the amplitudes of the FW  and BW wave, respectively.
Substituting Eq.~\eqref{E2} into Eq.~\eqref{WE1} and using slowly varying amplitudes approximation $k_2dE_2/dz\gg d^2E_2/dz^2 $ we obtain a system of two coupled equations for $A(z)$ and $B(z)$:
\begin{align}\label{QE}
\frac{dA}{dz}&=i\alpha A+i\sigma B e^{i2\Delta{k} z}, \nonumber\\
\frac{dB}{dz}&=-i\alpha B-i\sigma A e^{-i2\Delta{k} z},
\end{align}
where $\alpha={k_2\mu}/2$, $\sigma={k_2 p\mu}/4$, $\Delta{k}=k_1-k_2$.
The parameter $\sigma$ determines the coupling strength of two counterpropagating fields.

Linearly independent solutions  of Eqs.~\eqref{QE} are proportional to $\exp(\pm isz)$, where   $s=\sqrt{(\Delta k-\alpha)^2-\sigma^2}$. The probe field $E_2(z)$ in the layer is made of two counter-propagating waves $A(z)$  and $B(z)$. Their amplitudes are modulated with the spatial frequency $s$
\begin{align}
&A(z)=a_1e^{i\kappa_2^+z}+a_2e^{i\kappa_2^-z}=e^{ik_1z}(a_1e^{isz}+a_2e^{-isz}), \label{A}\\
&B(z)=b_1e^{-i\kappa_2^+z}+b_2e^{-i\kappa_2^-z}=e^{-ik_1z}(b_1e^{-isz}+b_2e^{isz}), \label{B}\\
&\kappa_2^\pm(\omega_2)=k_1\pm s.\label{k2} 
\end{align}
Equations \eqref{A} and \eqref{B} define the linearly independent solutions or normal waves (eigenwaves) in the approximation $|\Delta\varepsilon| \ll1$ and $|\Delta k|\ll|k_2|$. The normal waves are inhomogeneous: they are damped or amplified. We note that the forward and backward waves are a superposition of two spatial harmonics with wave vectors $\kappa_2^{\pm}$.  They may interfere with each other as follows from Eqs.~\eqref{A} and \eqref{B}. Expression \eqref{k2}  defines the dispersion relation.
Coefficients $a_{1,2}$ and $b_{1,2}$ are the constants of integration, determined by the boundary conditions.

Let us define the transmission and reflection coefficients for a layer of length $L$. Since $|\Delta\varepsilon|\ll1$,  we can neglect the Fresnel reflection from the layer boundaries and take into account only the volume reflection \cite{Karpov:PhUsp.36.1}.
Then the boundary conditions can be written as $A(0)=A_0$, $B(L)=0$, where  $A_0$  is the incident probe wave amplitude. By using these boundary conditions we find the integration constants:
\begin{align}\label{E4}
a_{1,2}&=A_0[s \mp (\Delta k-\alpha)]e^{\mp isL}/2D, \nonumber\\
b_{1,2}&=\pm iA_0\sigma e^{\pm isL}/2D,
\end{align}
where $D=s\cos(sL)+i(\Delta k-\alpha)\sin(sL)$.
We introduce the notation $t(z)=|A(z)|^2/|A_0|^2$  and $r(z)=|B(z)|^2/|A_0|^2$.
Using Eqs.~\eqref{A}--\eqref{E4}, one obtains
\begin{align}\label{E5}
  A(z)&=A_0[s\cos{s(L-z)}+i(\Delta{k}-\alpha)\sin{s(L-z)}]/D \nonumber\\
  B(z)&=A_0[\sigma \sin{s(L-z)}]/D.
\end{align}
The transmission $T$  and reflection $R$ coefficients  are given as
\begin{align}\label{E5a}
 T&=\frac{|A(L)|^2}{|A_0|^2}=\left|\frac{s}{s\cos(sL)+i(\Delta k-\alpha)\sin(sL)}\right|^2 \nonumber\\
 R&=\frac{|B(0)|^2}{|A_0|^2}=\left|\frac{\sigma \sin{sL}}{s\cos(sL)+i(\Delta k-\alpha)\sin(sL)}\right|^2.
\end{align}
\section{NUMERICAL RESULTS AND DISCUSSION }\label{resultsl}
\subsection{A. Raman induced grating ($\Omega_3=0$)}\label{results_A}
First consider the case when the control field is switched off ($\Omega_3=0$).
In this case  susceptibility (1) is given by
\begin{equation}\label{eq2}
\chi_{nl}(\omega_2)=\frac{1}{2} \chi_R |E_p|^2,
\end{equation}
where
\begin{equation}\label{chi_R}
\chi_R=\frac1{4\hbar^3} \frac{d_{21}^2d_{10}^2}{\delta_1 \delta_2\Delta_{20}}.
\end{equation}
Formula \eqref{chi_R} is the microscopic Raman susceptibility \cite{Boyd:NO}. We emphasize  that the imaginary part of $\chi_R$ is negative, i.e. the probe wave is amplified (negative absorption) due to energy transfer from the pump to the probe. The real part of $\chi_R$  has normal dispersion near the resonance, therefore the group velocity can be much less than the velocity of light in vacuum \cite{Payne:PRA.64.031802,Arkh:JETP.111.898}.

 For numerical simulations we use parameters corresponding to the $D_1$ line of Na atoms, and the levels $|0\rangle$ and $|2\rangle$ are long-lived hyperfine sublevels of the electronic ground state $3S_{1/2}$. The atomic parameters are $\gamma_{10}=2\pi\cdot10$~MHz, $\gamma_{20}=\gamma_{10}/100$, $N=10^{12}$~cm$^{-3}$, $\gamma_{10}=\gamma_{30}$.

Let us first consider a perfect standing pump field $E_1^+=E_1^-=E_1$ with the Rabi frequency of the pump field $\Omega_1=d_{01}E_1/\hbar$. In this case $|E_p(z)|^2$ varies periodically from $0$ to $4E_{1}^2$. The pump intensity vanishes at the nodal positions around which the atoms do not amplify the probe radiation because the Raman gain is zero. The probe propagates in the coherently dressed system described by Eq.~\eqref{WE} with Raman susceptibility \eqref{chi_R} as in a one-dimensional periodic system with a space period $\Lambda=\lambda_1/2$.

\begin{figure}
  \centering
   \includegraphics[width=.4\columnwidth]{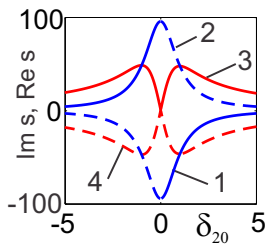}
  \caption{(Color online) Imaginary and real parts of $\pm s(\omega_2)=\kappa(\omega_2)-k_1$ as a function of the Raman  detuning $\delta_{20}$  for the Rabi frequency of the pump field $\Omega_1=1.5$. 1 -- $\Im(s)$, 2 -- $\Im(-s)$, 3 -- $\Re(s)$, 4 -- $\Re(-s)$. Here and below $\delta_{20}$ and $\Omega_1$ are in the units of $\gamma_{20}$ and $\gamma_{10}$, respectively.}\label{fig2}
\end{figure}

Typical real and imaginary parts of $\pm s$  are depicted in Fig.~\ref{fig2} as a function of the Raman detuning $\delta_{20}$.
Qualitative behavior of  $\Im s(\omega_2)$ and $\Re s(\omega_2)$ does not depend on the magnitude of the pump Rabi frequency. The only effect of the increased pump is quantitative changes.
Note that due to $\Delta k\neq 0$  the curves $\Re s(\omega_2)$  and $\Im s(\omega_2)$ are displaced slightly with respect to the resonance $\delta_{20}=0$ toward longer wavelengths.

\begin{figure} [!t]
  \centering
   \includegraphics[width=.75\columnwidth]{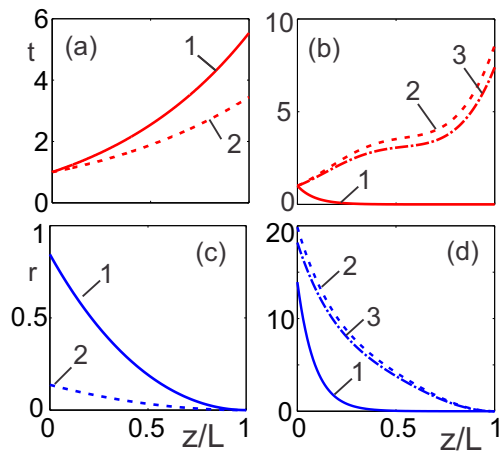}
  \caption{(Color online)
 Scaled intensity distribution of the forward $t(z/L)=|A(z/L)|^2/|A_0|^2$ (a,b) and backward $r(z/L)=|B(z)/L|^2/|A_0|^2$ (c,d) probe components inside the sample of length $L\simeq 0.6$ mm  for various Rabi frequencies of the pump field and Raman detuning.
(a) $\Omega_1=0.5\gamma_{10}$, ${1} - \delta_{20}=0$ and ${2}- \delta_{20}=\pm\gamma_{20}$;
(b) $\Omega_1=1.5\gamma_{10}$, ${1}- \delta_{20}=0$,  ${2}-\delta_{20}=-1.64\gamma_{20}$  and ${3}-\delta_{20}=1.64\gamma_{20}$; (c) $\Omega_1=0.5\gamma_{10}$, ${1} - \delta_{20}=0$ and ${2}- \delta_{20}=\pm\gamma_{20}$;
(d) $\Omega_1=1.5\gamma_{10}$, ${1}- \delta_{20}=0$, ${2}- \delta_{20}=-1.44\gamma_{20}$ and ${3}- \delta_{20}=1.44\gamma_{20}$.
}\label{fig2ac}
\end{figure}

Figure~\ref{fig2ac} shows the intensities of FW (a,b) and BW (c,d) probe waves in the sample as a function of the normalized coordinate $z/L$.  The behavior of the FW and BW waves depends on the Raman detuning and the pump intensity for the given length of a sample. It can be seen that they can either enhance or decay.

\begin{figure}[!b]
\centering
\includegraphics[width=.99\columnwidth]{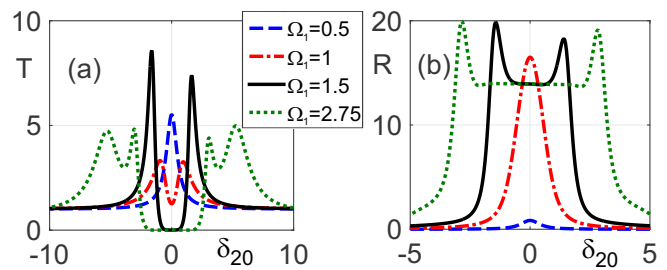}
\caption{(Color online) Transmission $T$ (a) and reflection $R$ (b) spectra vs  Raman detuning $\delta_{20}$ for various pump Rabi frequencies $\Omega_1$.
}\label{fig4}
\end{figure}

In Fig.~\ref{fig4}, the transmittance $T$ and reflectivity $R$ for the probe are plotted as functions of the Raman detuning for different  $\Omega_1$. We fix the one-photon detuning of the pump field ($\delta_{1}=30\gamma_{10}$) and tune $\delta_2$.
It can be seen that the transmission and reflection  spectra strongly depend on the pump intensity.
The transmitted and reflected light can be simultaneously amplified   in some frequency range depending on the pump intensity. Therefore,  transmittance can be interpreted as a transmission gain or  gain spectrum.
We note that near the Raman resonance $T\to 0$ and a transparent medium becomes opaque.
This structure can also be considered as a nonlinear mirror with the reflectivity $R>1$
near the Raman resonance (Fig.~\ref{fig4}b).

\begin{figure}[!h]
  \centering
   \includegraphics[width=.95\columnwidth]{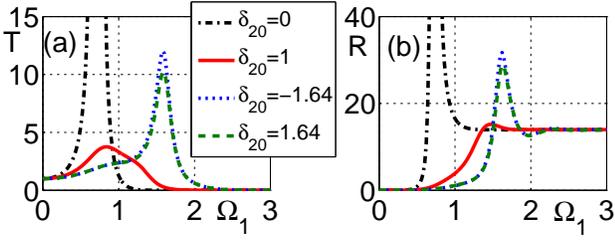}
     \caption{(Color online) Transmission $T$ (a) and  reflectivity $R$ (b) as a function of the pump Rabi frequency $\Omega_1$ for various  Raman detunings $\delta_{20}$.
     } \label{fig5}
\end{figure}

In Fig.~\ref{fig5}, the transmittance and reflectivity are plotted as a function of the pump Rabi frequency for different Raman detunings $\delta_{20}$. In the case of resonance $\delta_{20}=0$  the transmittance and reflectivity  grow as the pump intensity increases reaching their maximum at some value of $\Omega_1^m$.
For $\Omega_1>\Omega_1^m$  the transmission and reflection are reduced and the transmission and reflection  spectra  split into a doublet (see Fig.~\ref{fig4}) and a dip  occurs in the vicinity of the Raman resonance. The dip width and  depth in the transmission spectrum   increase with the pump field and additional peaks emerge  (Fig.~\ref{fig4}a). The distance between the peaks in the reflection spectrum also increases with the pump field, but the depth of the dip does not change (Fig.~\ref{fig4}b). The asymmetry of the peaks is observed in transmission and reflection  due to $\Delta k\neq 0$. Away from the Raman resonance  $T\to 1$ and  $R\to 0$.
When $\delta_{20}\neq0$, the maximum in transmission is reduced and occurs at higher pump fields (Fig.~\ref{fig5}).

\begin{figure}[!h]
\centering
\includegraphics[width=.95\columnwidth]{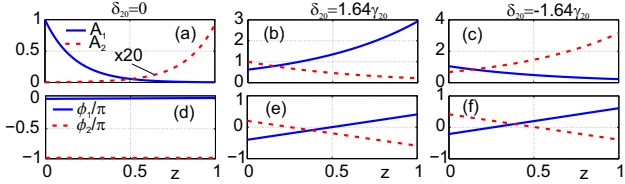}
\caption{(Color online) Spatial distribution of the amplitude $A_{1,2}(z)/A_0$ $(a-c)$ and  phase $\varphi_{1,2}/\pi$ $(d-f)$ of spatial harmonics for the FW probe wave inside the sample for frequency detunings corresponding to the Raman resonance $\delta_{20}=0$ and transmission peaks $\delta_{20}=\pm 1.64\gamma_{10}$ when $\Omega_{1}=1.5\gamma_{10}$.
} \label{fig6}
\end{figure}

\begin{figure}[!h]
\centering
\includegraphics[width=.6\columnwidth]{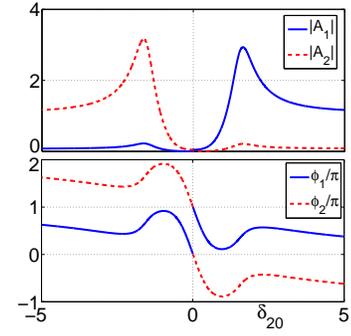}
\caption{(Color online)
The amplitudes $A_{1,2}(L)$ and  phases $\varphi_{1,2}(L)$  vs  Raman deturning $\delta_{20}$ at the output of the sample under $\Omega_1=1.5\gamma_{10}$.} \label{fig6-1}
\end{figure}

This transmission and reflection behavior is due to the fact that FW and BW probe waves are a superposition of two spatial harmonics (see Eqs.~\eqref{A} and \eqref{B}).
In Fig.~\ref{fig6}(a-f), the  amplitudes  $A_{1,2}(z)=|a_{1,2}e^{\pm isz}|$ $(a-c)$ and phases $\varphi_{1,2}$ $(d-f)$ of the spatial harmonics for the FW wave $E_2^+$ are plotted as a function of the  coordinate $z/L$ inside a sample. It is seen that one of the amplitudes increases while the other one decreases, and they may be in phase or in antiphase.
The total field is a result of interference between harmonics. In a resonance region, harmonic amplitudes are close and they are in antiphase (see Fig.~\ref{fig6-1}), therefore the transmission is close to zero (destructive interference). Clearly, the amplifying harmonic gives the main contribution to the transmission maximum. The amplified and damped harmonics  interchange when passing through the Raman resonance.

\begin{figure}[!h]
\centering
\includegraphics[width=.6\columnwidth]{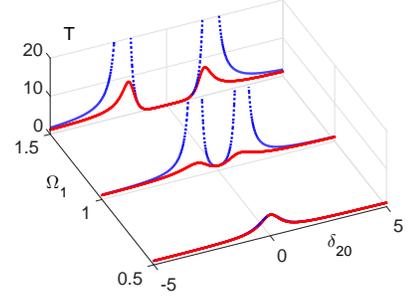}
\caption{(Color online) The transmission spectra of RIG for different $\Omega_1$. The blue line is the transmission spectra formed only by the Raman grating when $\Re\chi_R=0$ (see Eq. \eqref{chi_R}) and
the red line is the spectra formed by both gratings} \label{fig6-2}
\end{figure}

Transmission and reflection spectra are formed by a joint action of the gain grating and the refraction grating. The latter occurs when $\delta_{20}\neq0$. From Fig.~\ref{fig6-2} we can see that under the Raman resonance the transmission is determined only by the gain grating, while out off resonance the spectrum is determined by  both gratings.

\begin{figure}[!h]
  \centering
   \includegraphics[width=.9\columnwidth]{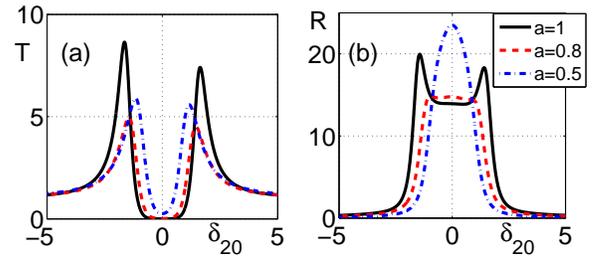}
   \caption{(Color online) The transmission $T$ (a) and  reflectivity $R$ (b) spectra vs  Raman detuning $\delta_{20}$ for various values of the parameter $a$.
   $\Omega_1^+=1.5\gamma_{10}$} \label{fig7}
\end{figure}

Generally, $E_1^+\neq E_1^-$ and the standing wave is not perfect. In this case the pump intensity does not vanish anywhere with the nodes becoming quasi-nodes and Raman amplification taking place in the quasi-nodes. Formulas for $\Delta \varepsilon$ and $p$ in Eq.~\eqref{WE} are conveniently represented as $\Delta\varepsilon=4\pi\chi_R(\omega_2)NE_{1+}^2(1+a^2)$ and $p=2a/(1+a^2)$, where $a=E_1^-/E_1^+=\Omega_1^-/\Omega_1^+$.
Fig.~\ref{fig7} shows the transmission and reflection spectra for different values of the parameter $a$.
Here, the solid curves correspond to the ideal pump standing wave ($a=1$). It is seen that these spectra are not very sensitive to the value of $a$.
The transmission and reflection dependences as a function of the pump Rabi frequency are the same as for $a=1$, but the curves are shifted to higher pump fields since  the modulation depth  of the grating is reduced.

\subsection{B. Coherent control of a Raman induced grating ($\Omega_3\neq0$) }\label{results_B}

Now we consider an ARG scheme (Fig.~\ref{fig1}) when a control field is switched on ($\Omega_3\neq0$). In this case, the nonlinear Raman susceptibility is determined by Eqs.~\eqref{eq1}.
For a weak control field,  the gain profile exhibits a maximum near resonance ($\delta_{20}=0$). However, the gain peak is split into a doublet, i.e. the dip in line center of the gain profile appears \cite{Agarwal:PRA.70.023802,Arkh:PRA.88.033847} for $\Omega_3$ exceeding some critical value $\Omega_3^{cr}$.
The width and depth of the dip increase with $\Omega_3$. The appearance of a gain-doublet structure   leads to anomalous dispersion $\Re[\chi_{nl}(\omega_2)]$  near the Raman resonance.

\begin{figure}[!h]
  \centering
   \includegraphics[width=.95\columnwidth]{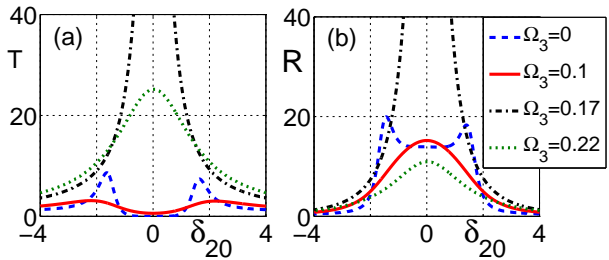}
   \caption{(Color online) The transmission $T$ ($a$) and reflection $R$ (b) spectra  as a function of the Raman detuning $\delta_{20}$  for various Rabi frequencies of the control field  $\Omega_3$ (in units of $\gamma_{10}$).
   $\Omega_1=1.5\gamma_{10}$, $a=1$, $\delta_3=0$ }\label{fig8}
\end{figure}

Let us show that using a weak additional coherent field ($\Omega_3<\Omega_3^{cr}$), we can control the spectral properties of the RIG.
Figure~\ref{fig8} shows  transmission (a)  and  reflection (b) of the probe field
as functions of the Raman detuning $\delta_{20}$ for different Rabi frequencies of the control field being in resonance  with the corresponding transition $|2\rangle$-$|3\rangle$. Initially, when $\Omega_3=0$, the pump intensity is chosen such that the transmission is close to zero at the $\delta_{20}=0$. As can be seen from Fig.~\ref{fig8}, the transmission and reflection spectra critically  depend on the control field intensity. Small variations in the control field  intensity  can change the system from opaque to transparent state (with amplification) and vice versa. Thus, this structure can operate as an all-optical transistor.

\begin{figure}[!t]
  \centering
   \includegraphics[width=.95\columnwidth]{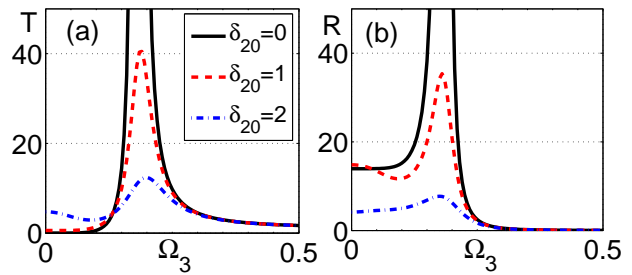}
   \caption{(Color online) Transmission $T$ ($a$) and reflection $R$ (b) as a function of the pump Rabi frequency for different Raman detunings $\delta_{20}$. $\Omega_1^+=1.5\gamma_{10}$, $a=1$, $\delta_3=0$ }\label{fig9}
\end{figure}

\begin{figure}[!t ]
  \centering
   \includegraphics[width=.9\columnwidth]{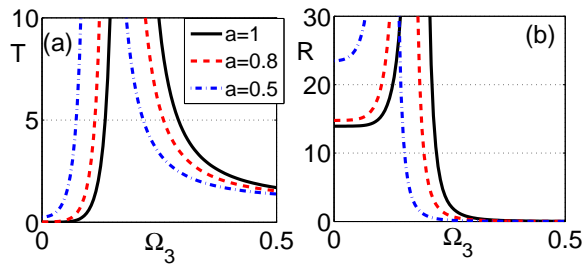}
   \caption{(Color online) Transmission $T$ ($a$) and reflection $R$ (b) as a function of the control field Rabi frequency $\Omega_3$ for different values of the parameter $a$.
    $\Omega_1^+=1.5\gamma_{10}$, $\delta_{20}=\delta_3=0$}\label{fig11}
\end{figure}

Transmission and reflection is strongly dependent on the control field Rabi frequency  and  Raman detuning (Fig.~\ref{fig9}). At first, as $\Omega_3$ increases up to some value so does  the transmission until it reaches its  maximum and then drops. The larger the Raman detuning, the lower the transmission maximum. Reflection behaves similarly (Fig.~\ref{fig9}b).
In the case of a non-ideal standing pump wave ($a\neq 0$) the transmission and reflection curves as functions of $\Omega_3$ are displaced to smaller  control fields, as is seen from Fig.~\ref{fig11}.

\section{CONCLUSIONS}

We have theoretically investigated  propagation of a probe field in  three- and four-level atomic media under Raman interaction with a standing-wave pump field. The probe field has been found to  undergo a periodic variation of the Raman gain and refractive index due to periodic spatial modulation of the permittivity.
Therefore the medium acts as a gain and refraction grating (a hybrid grating), and can dramatically  change the spectral and  transmission properties of the medium. These gratings  are fundamentally different from the recent  EIAG schemes.

It has been  shown that a transparent medium with a periodic spatial modulation of the Raman gain can be opaque near the Raman resonance. This conclusion is of a counterintuitive character. At the same time, it represents a non-linear  selective mirror with $R>1$.
In some spectral range, the transmission and reflection can be amplified simultaneously and   dynamically tuned by varying the intensity of the pump field. The pump wave intensity required for these effects to be observed depends on a number of factors (one-photon pump frequency detuning, Raman resonance width and the length of medium) and can be anything from 10 to $100$~mW/cm$^2$ and less.

We also have shown that the spectral properties of a Raman induced grating can be controlled by using of an additional weak control field. Small variations in the resonant control field intensity can change the system from opaque  to  transparent. Such controllable transmission is well suited to study  all-optical switching at low control field intensities  and create an all-optical transistor.
The control field intensity can be anything from 10 to 100 mkW/cm$^2$.

For experimental realization both room-temperature and ultracold atoms or ions as well as  molecular gases can be used. These experiments are similar to  \cite{Brown:OL.30.699}. Hollow-core photonic crystal fibers filled by atoms can be used to reduce the pump and control field intensity requirements.

The results obtained are in good agreement with exact calculations based on  the recurrence-relations technique  \cite{Arkh:QE.39.157}.

This work was supported  by the RFBR  through Grant 15-02-03959.


%

\end{document}